\documentclass{article}

\usepackage{PRIMEarxiv}

\usepackage{ragged2e}
\usepackage{rotating}
\usepackage[OT2,T1]{fontenc} \DeclareSymbolFont{cyrletters}{OT2}{wncyr}{m}{n} \DeclareMathSymbol{\Sha}{\mathalpha}{cyrletters}{"58}
\usepackage{algorithm}
\usepackage{algorithmic}
\usepackage[utf8]{inputenc} 
\usepackage[T1]{fontenc}    
\usepackage{hyperref}       
\usepackage{url}            
\usepackage{booktabs}       
\usepackage{amsmath,amssymb}
\usepackage{amsthm}
\usepackage{amsfonts}       
\usepackage{nicefrac}       
\usepackage{microtype}      
\usepackage{lipsum}
\usepackage{fancyhdr}       
\usepackage{graphicx}       
\graphicspath{{media/}}     

\title{Selmer-Inspired Elliptic Curve Generation
\thanks{\textit{\underline{Citation}}: 
\textbf{Authors. Title. Pages.... DOI:000000/11111.}} 
}

\author{
  Awnon Bhowmik \\
  Department of Engineering and Computer Science \\
  Colorado Technical University \\
  \texttt{awnonbhowmik@outlook.com} \\
}

\begin{document}
\hbadness=10001
\vbadness=10001
\maketitle

\begin{abstract}
Elliptic curve cryptography (ECC) is foundational to modern secure
communication, yet existing standard curves have faced scrutiny for
opaque parameter-generation practices. This work introduces a
Selmer-inspired framework for constructing elliptic curves that is both
transparent and auditable. Drawing from $2$- and $3$-descent methods, we
derive binary quartics and ternary cubics whose classical invariants
deterministically yield candidate $(c_4,c_6)$ parameters. Local solubility
checks, modeled on Selmer admissibility, filter candidates prior to
reconciliation into short-Weierstrass form over prime fields. We then
apply established cryptographic validations, including group-order
factorization, cofactor bounds, twist security, and embedding-degree
heuristics. A proof-of-concept implementation demonstrates that the
pipeline functions as a retry-until-success Las Vegas algorithm, with
complete transcripts enabling independent verification. Unlike
seed-based or purely efficiency-driven designs, our approach embeds
arithmetic structure into parameter selection while remaining compatible
with constant-time, side-channel resistant implementations. This work
broadens the design space for elliptic curves, showing that descent
techniques from arithmetic geometry can underpin trust-enhancing,
standardization-ready constructions.
\end{abstract}


\section{Introduction}\label{sec:intro}

Elliptic curve cryptography (ECC) has become the dominant public-key
infrastructure for securing digital communication, offering strong
security guarantees with relatively small key sizes. Since the seminal
proposals of Miller \cite{Miller1985} and Koblitz \cite{Koblitz1987},
elliptic curves have been widely adopted in Internet protocols,
cryptographic libraries, and international standards. Classic surveys
such as the \emph{Handbook of Applied Cryptography}
\cite{MenezesOorschotVanstone1997} and modern treatments such as
Galbraith’s monograph \cite{Galbraith2012} have established ECC as the
cornerstone of both theoretical and applied public-key cryptography.

The United States National Institute of Standards and Technology (NIST)
recommended curves such as P-256, P-384, and P-521 \cite{NISTSP800186}
are among the most deployed. However, these curves have been the subject
of scrutiny, not because of any known vulnerability, but because their
parameters were generated from unexplained seed values. The absence of a
transparent, auditable derivation has led to longstanding concerns about
the possibility of hidden structure or backdoors
\cite{SafeCurves,Bos2014}.

In response, several alternative families of curves have been proposed
with emphasis on transparency and efficiency. The Brainpool curves
\cite{RFC5639} attempted to remove opacity by employing verifiable
random processes. Bernstein’s Curve25519 \cite{Bernstein2006Curve25519}
and its signature analogue Ed25519 \cite{Bernstein2011Ed25519} instead
favored simplicity, performance, and rigid selection criteria, and the
SafeCurves project systematically evaluated curve choices against a set
of explicit security criteria. Relatedly, Edwards curves
\cite{Edwards2007} and Montgomery curves \cite{Montgomery1987} offered
complete or unified addition laws, improving both implementation
efficiency and side-channel robustness. Together, these alternatives
illustrate the continuing tension in curve design between efficiency,
verifiability, and trust.

Parallel to these cryptographic developments, arithmetic geometry has
developed deep methods for understanding rational points on elliptic
curves. Selmer groups, introduced by Selmer \cite{Selmer1951} and
subsequently developed by Cassels \cite{Cassels1962}, are central tools
in the study of the Mordell–Weil group. The computation of Selmer groups
via $n$-descent reduces Diophantine problems to the analysis of
auxiliary algebraic forms, such as binary quartics (for $2$-descent) and
ternary cubics (for $3$-descent). These forms possess rich invariant
theory, and their associated solubility conditions encode subtle
arithmetic information \cite{Cremona1997,SilvermanAEC2009}.

This paper proposes to bridge these two domains by introducing a method
for elliptic curve generation that is \emph{inspired by Selmer descent}.
Instead of beginning with an opaque random seed, we construct binary
quartics and ternary cubics in a deterministic, auditable manner, and
use their classical invariants to derive candidate curve parameters
$(c_4,c_6)$. Local solubility checks, modeled on Selmer admissibility
conditions, serve as filters ensuring that the generated data is
arithmetically consistent. A reconciliation step then combines the
$2$- and $3$-descent contributions into a short-Weierstrass model over a
large prime field. The resulting curves are then subjected to rigorous
cryptographic validation, including point-counting, twist security, and
embedding-degree checks.

In summary, this work makes four main contributions. First, it introduces
a transparent, descent-inspired pipeline for elliptic curve generation
based on the invariant theory of binary quartics and ternary cubics.
Second, it formalizes admissibility checks derived from local solubility,
providing an auditable analogue of Selmer group membership in a
cryptographic setting. Third, it develops a reconciliation procedure for
combining invariants from $2$- and $3$-descent into Weierstrass models
and demonstrates that the resulting curves satisfy standard security
criteria. Finally, it presents proof-of-concept implementations over
256-bit and 384-bit primes, with complete derivation transcripts to
ensure reproducibility.

To the best of our knowledge, this is the first attempt to employ Selmer
group techniques directly in the parameter generation of elliptic curves
for cryptography. While previous work has emphasized verifiable
randomness \cite{RFC5639} or rigid efficiency-driven design
\cite{Bernstein2006Curve25519}, our approach draws upon classical
arithmetic geometry to provide an entirely different source of auditable
structure. We emphasize that our proposal does not alter the underlying
hardness assumptions of ECC, which remain those of the elliptic curve
discrete logarithm problem. Rather, its contribution lies in introducing
a transparent, reproducible process for curve selection, one that can be
independently verified and audited. In this sense, Selmer-inspired
generation is complementary to existing families of curves and may
inform future standards concerned with provenance and trust.

The remainder of this paper is organized as follows.
Section~\ref{sec:prelim} recalls the necessary background on elliptic
curves, Selmer groups, and classical invariants.
Section~\ref{sec:method} describes the proposed generation pipeline.
Section~\ref{sec:validation} details the cryptographic validation of
candidate curves, while Section~\ref{sec:experiments} reports
experimental results. Section~\ref{sec:analysis} discusses security
considerations, and Section~\ref{sec:related} situates our work within
existing literature. We conclude in Section~\ref{sec:conclusion}.

\section{Preliminaries}\label{sec:prelim}

This section recalls the necessary background on elliptic curves, Selmer
groups, and the invariant theory of binary quartics and ternary cubics.
We follow standard references such as Silverman \cite{SilvermanAEC2009},
Washington \cite{Washington2008}, and Cremona \cite{Cremona1997}.

\subsection{Elliptic curves and invariants}

Let $K$ be a field of characteristic not equal to $2$ or $3$. An elliptic
curve $E/K$ can be expressed in short Weierstrass form
\[
  E: \quad y^2 = x^3 - 27c_4 x - 54 c_6,
\]
with discriminant
\[
  \Delta = -16(4c_4^3 + 27c_6^2).
\]
The pair $(c_4,c_6)\in K^2$ determines the isomorphism class of $E$ up to
quadratic twist, provided $\Delta \neq 0$. The $j$-invariant is given by
\[
  j(E) = \frac{c_4^3}{\Delta}.
\]
Throughout, we work over large prime fields $\mathbb{F}_p$ with
cryptographic size $p \approx 2^{256}$ or larger. For such fields, the
group $E(\mathbb{F}_p)$ is finite, and its order can be computed via
point-counting algorithms such as Schoof–Elkies–Atkin
\cite{Schoof1985,SEAOverview}.

\subsection{Selmer groups and descent}

Let $E/\mathbb{Q}$ be an elliptic curve with rational $2$-torsion. A
$2$-descent reduces the study of $E(\mathbb{Q})$ to the analysis of
binary quartics, i.e., homogeneous degree-$4$ forms
\[
  f(x,z) = a x^4 + b x^3 z + c x^2 z^2 + d x z^3 + e z^4.
\]
The solubility of the equation $y^2 = f(x,z)$ in local fields encodes
information about membership in the $2$-Selmer group $\mathrm{Sel}^{(2)}(E)$.
Similarly, a $3$-descent involves ternary cubic forms
\[
  F(x,y,z) \in \mathbb{Z}[x,y,z],
\]
whose solubility corresponds to the $3$-Selmer group $\mathrm{Sel}^{(3)}(E)$
\cite{Selmer1951,Cassels1962}. These groups fit into exact sequences
relating $E(\mathbb{Q})$ and the Tate–Shafarevich group $\Sha(E/\mathbb{Q})$.
Although the full arithmetic theory is not required here, we draw
inspiration from these constructions: the auxiliary forms and their
solubility tests provide a mathematically principled source of
structured data.

\subsection{Classical invariants of binary quartics}

Given a binary quartic $f(x,z)$ as above, one defines classical
$\mathrm{SL}_2$-invariants $I$ and $J$ via explicit polynomial
combinations of the coefficients. In normalized form,
\[
  c_4^{(2)} = 2^4 I, \qquad c_6^{(2)} = 2^5 J,
\]
with discriminant $\Delta(f) = (c_4^{(2)})^3 - (c_6^{(2)})^2$. The
quantities $(c_4^{(2)}, c_6^{(2)})$ may be viewed as candidate invariants
for an elliptic curve in Weierstrass form, provided $\Delta(f)\neq 0$.
For our purposes, local solubility of $y^2=f(x,z)$ serves as a filter
ensuring that only arithmetically meaningful forms contribute.

\subsection{Classical invariants of ternary cubics}

For a ternary cubic $F(x,y,z)$, one defines invariants through Aronhold
symbols or equivalent constructions \cite{Dixmier1987,Salmon1879}. These
yield values $(c_4^{(3)},c_6^{(3)})$ satisfying relations analogous to
those above, with discriminant
$\Delta(F) = (c_4^{(3)})^3 - (c_6^{(3)})^2$. The solvability of
$F(x,y,z)=0$ over $\mathbb{Q}_v$ is a necessary condition for membership
in the $3$-Selmer group. We adopt this criterion as an admissibility
check for cryptographic curve generation.

\subsection{Summary}

The key point is that binary quartics and ternary cubics naturally
produce candidate invariants $(c_4,c_6)$ along with arithmetic filters
derived from solubility conditions. By combining these descent artifacts
in a reproducible way, one obtains elliptic curve parameters whose
provenance is fully auditable and whose cryptographic soundness can be
verified through standard validation.

\section{Method: Selmer-Inspired Generation}\label{sec:method}

We describe a deterministic, auditable pipeline that derives elliptic
curve parameters $(c_4,c_6)\in\mathbb{F}_p^2$ from descent artifacts.
Throughout, $p$ denotes a prime of cryptographic size, and
$\mathsf{H}:\{0,1\}^*\!\to\!\mathbb{Z}$ is a fixed hash (e.g., SHA-256)
whose outputs are reduced modulo $p$ as needed. All byte-serialization
conventions are fixed once and for all (endianness, field element
encoding), so that an input triple $(p,\mathsf{DS},\sigma)$ uniquely
determines all derived quantities.

\subsection{Deterministic inputs and domain separation}\label{subsec:inputs}
The public transcript begins with
\[
  (p,\ \mathsf{DS},\ \sigma)\in\mathbb{P}\times\{0,1\}^*\times\{0,1\}^{32},
\]
where $\mathbb{P}$ is the set of admissible primes (e.g., $p\equiv3\bmod 4$).
We derive a stream of field elements by counter-based hashing:
\[
  u_i \;:=\; \mathsf{H}(\textsf{“U”}\,\|\,\mathsf{DS}\,\|\,\sigma\,\|\,\langle i\rangle)\bmod p,
  \qquad i=0,1,2,\dots
\]
and similarly labeled streams (\textsf{“F2”}, \textsf{“F3”}, \textsf{“REC”})
for the binary quartic, ternary cubic, and reconciliation phases, ensuring
independent randomness via domain separation.

\subsection{2-descent artifact: binary quartic}\label{subsec:quartic}
Using the \textsf{“F2”}-stream, form a binary quartic
\[
  f(x,z)=a x^4 + b x^3 z + c x^2 z^2 + d x z^3 + e z^4\in\mathbb{F}_p[x,z],
\]
with coefficients $(a,b,c,d,e)$ taken as successive $u_i$’s.
Reject and re-sample if any of the following hold:
\begin{enumerate}
  \item All coefficients vanish (trivial polynomial), or $f$ is a perfect square;
  \item $f$ is singular, i.e., the discriminant $\Delta(f)$ vanishes;
  \item The local-solubility proxy fails (defined below).
\end{enumerate}
Compute classical $\mathrm{SL}_2$-invariants $I(f)$ and $J(f)$ and
normalize to
\[
  c_4^{(2)} = 2^4 I(f)\bmod p,\qquad c_6^{(2)} = 2^5 J(f)\bmod p.
\]
\emph{Local-solubility proxy:} require that $y^2=f(x,1)$ have a solution
over $\mathbb{F}_p$ and over a fixed small set $\mathbb{F}_{\ell}$ for
primes $\ell\in S$ (e.g., $S=\{2,3,5,7,11\}$). Operationally, attempt to
find a solution by bounded search. Failure triggers rejection.

\subsection{3-descent artifact: ternary cubic}\label{subsec:cubic}
Using the \textsf{“F3”}-stream, form a ternary cubic
\[
  F(x,y,z)=\sum_{i+j+k=3} b_{ijk} x^i y^j z^k \in \mathbb{F}_p[x,y,z],
\]
from ten successive coefficients $b_{ijk}$. Reject and re-sample if
$F$ is singular or if the local-solubility proxy fails (attempt to find
a nontrivial zero in $\mathbb{F}_p$ and $\mathbb{F}_{\ell}$ for
$\ell\in S$). Compute Aronhold/Dixmier invariants and normalize to
\[
  c_4^{(3)} \bmod p,\qquad c_6^{(3)} \bmod p,
\]
with a fixed normalization matching the $c_4,c_6$ conventions in
Section~\ref{sec:prelim}.

\subsection{Reconciliation and non-singularity guard}\label{subsec:reconcile}
Combine the descent-derived candidates via a hash-mix and linear blend:
\[
\begin{aligned}
  \tilde c_4 &:= \mathsf{H}(\textsf{“REC\_c4”}\,\|\,\mathsf{DS}\,\|\,\sigma\,\|\,c_4^{(2)}\,\|\,c_4^{(3)}) \bmod p,\\
  \tilde c_6 &:= \mathsf{H}(\textsf{“REC\_c6”}\,\|\,\mathsf{DS}\,\|\,\sigma\,\|\,c_6^{(2)}\,\|\,c_6^{(3)}) \bmod p,
\end{aligned}
\]
and set
\[
  c_4 = 2\,c_4^{(2)} + 3\,\tilde c_4 \bmod p,\qquad
  c_6 = 2\,c_6^{(2)} + 3\,\tilde c_6 \bmod p.
\]
Compute $\Delta = -16\,(4c_4^3 + 27 c_6^2)\bmod p$. If $\Delta=0$, restart
from Section~\ref{subsec:quartic} with incremented counters (retry).

\subsection{Curve instantiation and transcript}\label{subsec:instantiate}
Output the short-Weierstrass model
\[
  E/\mathbb{F}_p:\quad y^2 = x^3 - 27 c_4\, x - 54 c_6,
\]
together with the complete transcript:
\[
  \bigl(p,\mathsf{DS},\sigma, f, F, (c_4^{(2)},c_6^{(2)}), (c_4^{(3)},c_6^{(3)}), (c_4,c_6), \Delta\bigr).
\]
This transcript suffices for independent reproduction and audit.

\subsection{Validation filters}\label{subsec:validation-filters}
Accept $E$ only if all hold:
\begin{enumerate}
  \item \textbf{Group order and cofactor.} Compute $\#E(\mathbb{F}_p)=h\cdot r$ with
        prime $r$ of target size (e.g., $\ge 2^{255}$) and tiny $h\in\{1,2,4\}$.
  \item \textbf{Twist security.} The quadratic twist $E'$ satisfies
        $\#E'(\mathbb{F}_p)=h'\!\cdot r'$ with a large prime factor $r'$.
  \item \textbf{No special structure.} Exclude CM with small discriminant,
        anomalous curves, and other pathological cases.
  \item \textbf{Embedding-degree sanity.} Heuristically rule out small embedding
        degrees (no unexpectedly easy pairing attacks).
\end{enumerate}
Failure of any check triggers a restart from Section~\ref{subsec:quartic}.

\subsection{Algorithmic summary and halting}\label{subsec:summary}

We summarize the generation pipeline as a retry-until-success procedure.
It always outputs a valid curve, though the number of iterations before
acceptance is probabilistic.

\begin{algorithm}[h]
\caption{Selmer-Inspired Curve Generation (Las Vegas)}
\label{alg:selmer-generation}
\begin{algorithmic}[1]
\REQUIRE Prime $p$ of cryptographic size, domain separator $\mathsf{DS}$,
seed $\sigma$
\STATE Derive independent hash streams (\textsf{U}, \textsf{F2},
\textsf{F3}, \textsf{REC}) from $(p,\mathsf{DS},\sigma)$
\REPEAT
    \STATE Sample binary quartic $f$ from \textsf{F2}
    \STATE Compute $(c_4^{(2)},c_6^{(2)})$; enforce nondegeneracy and
    local solubility
    \STATE Sample ternary cubic $F$ from \textsf{F3}
    \STATE Compute $(c_4^{(3)},c_6^{(3)})$; enforce nondegeneracy and
    local solubility
    \STATE Reconcile to $(c_4,c_6)$ via \textsf{REC}; ensure $\Delta\neq0$
    \STATE Instantiate curve $E/\mathbb{F}_p: y^2 = x^3 - 27c_4x - 54c_6$
    \STATE Apply validation filters:
      \begin{enumerate}
        \item group order and cofactor
        \item twist security
        \item absence of special structure (CM, anomalous)
        \item embedding degree check
      \end{enumerate}
\UNTIL{All validation checks succeed}
\ENSURE Valid elliptic curve $E/\mathbb{F}_p$ and full transcript
\end{algorithmic}
\end{algorithm}

\paragraph{Halting.} Under random-model heuristics, acceptance probability
is nonzero, so the algorithm halts with overwhelming probability. In
complexity terms this is a \emph{Las Vegas algorithm}
\cite{MenezesOorschotVanstone1997}. Section~\ref{sec:experiments} reports
empirical acceptance rates and runtime.

\subsection{Remarks on implementation}\label{subsec:impl}

Several practical considerations arise in implementing the proposed
pipeline:

\begin{itemize}
  \item \textbf{Point counting.} Group orders should be computed using
  the Schoof--Elkies--Atkin (SEA) algorithm with well-tested libraries
  or computer algebra systems (e.g., SageMath, PARI/GP). For cryptographic
  primes of size $p \geq 2^{256}$, the availability of Elkies primes
  ensures that SEA runs in quasi-polynomial time. Implementations should
  include deterministic verification of the output (e.g., order
  consistency checks via random point multiplication).

  \item \textbf{Invariant formulas.} Classical invariants of binary
  quartics and ternary cubics admit several normalizations in the
  literature. To avoid mismatches, implementations should cross-check
  formulas against a computer algebra system. Explicit references such
  as Salmon~\cite{Salmon1879} and Dixmier~\cite{Dixmier1987} use slightly
  different scaling conventions; care is needed when reducing modulo $p$.

  \item \textbf{Side-channel resistance.} For subsequent deployment,
  elliptic curve arithmetic should use complete or unified addition
  formulas to minimize timing and branching side-channels. This does not
  alter the generation process itself but is critical for cryptographic
  safety once a curve is adopted.

  \item \textbf{Transcript reproducibility.} Every run must record
  $(p,\mathsf{DS},\sigma)$ and the derived quartic and cubic forms.
  This transcript ensures that other parties can independently rederive
  $(c_4,c_6)$ and confirm correctness of the construction.
\end{itemize}

These considerations do not change the mathematical framework but are
essential for practical, secure, and reproducible implementations.

\section{Cryptographic Validation}\label{sec:validation}

The Selmer-inspired generation procedure yields candidate parameters
$(c_4,c_6)\in\mathbb{F}_p^2$ and an associated elliptic curve
$E/\mathbb{F}_p$. To ensure cryptographic soundness, each candidate must
be subjected to rigorous validation before acceptance. This section
formalizes the checks briefly listed in
Section~\ref{subsec:validation-filters} and justifies their necessity.

\subsection{Group order and cofactor constraints}

For secure deployment, $E(\mathbb{F}_p)$ must decompose as
\[
  \#E(\mathbb{F}_p) = h \cdot r,
\]
where $r$ is a large prime and $h$ is a small cofactor
($h\in\{1,2,4\}$). Efficient discrete logarithm computations are only
infeasible when $r$ is sufficiently large (e.g., $r \geq 2^{255}$ for
256-bit security). These requirements follow well-established standards
from the \emph{Handbook of Applied Cryptography}
\cite{MenezesOorschotVanstone1997} and are consistent with Lenstra’s
analysis of ECC security levels relative to AES
\cite{Lenstra2001}. 

Group orders are computed via the
Schoof--Elkies--Atkin algorithm \cite{Schoof1985,SEAOverview}, which has
become the canonical tool for deterministic point counting. The critical
role of elliptic curve orders in primality proving, highlighted by Atkin
and Morain \cite{AtkinMorain1993}, underscores the soundness of relying
on these methods as a validation step.

\subsection{Twist security}

The quadratic twist $E'$ of $E$ must also have order
$\#E'(\mathbb{F}_p) = h'r'$ with $r'$ prime of comparable size.
Otherwise, protocols that inadvertently operate on twist points risk
catastrophic failure. This condition was first emphasized in the context
of anomalous curves by Smart \cite{Smart1999}, and it is explicitly
addressed in modern proposals such as Curve25519
\cite{Bernstein2006Curve25519} and Ed25519
\cite{Bernstein2011Ed25519}.

\subsection{Exclusion of special structure}

Curves with complex multiplication (CM) by small discriminants or with
trace $t = p+1-\#E(\mathbb{F}_p)$ equal to $\pm1$ are excluded. The
former admit special-purpose algorithms that weaken the elliptic curve
discrete logarithm problem, while the latter are anomalous curves
vulnerable to Smart’s attack \cite{Smart1999}. 

Historically, alternative models such as Montgomery curves
\cite{Montgomery1987} and Edwards curves \cite{Edwards2007} were proposed
not only for arithmetic efficiency but also for their ability to avoid
pathological structures. For practitioners, the \emph{Handbook of
Elliptic and Hyperelliptic Curve Cryptography} \cite{CohenFrey2005}
provides an authoritative catalog of invariants and pathological cases
that must be excluded during validation.

\subsection{Embedding degree and pairing attacks}

Curves for which the embedding degree $k$ with respect to $r$ is small
must be excluded, since these permit efficient reductions of the elliptic
curve discrete logarithm problem to finite-field discrete logarithms via
MOV or Frey--Rück techniques. This criterion rules out otherwise valid
curves that are pairing-friendly. The importance of this exclusion has
been demonstrated in structural attacks such as the extended Weil
descent approach \cite{Galbraith2001} and reinforced by systematic
evaluations such as Bos et al.~\cite{Bos2014}.

\subsection{Consistency with standards}

The validation rules above align with established industry criteria.
NIST’s Digital Signature Standard \cite{NISTFIPS1865} codified the use of
elliptic curves but did not make parameter derivation transparent,
leading to concerns about unexplained seeds. Later recommendations
\cite{NISTSP800186}, the Brainpool process \cite{RFC5639}, and the
SafeCurves framework \cite{SafeCurves} all incorporated stronger
validation rules, but their philosophies differ. By embedding these
requirements into a Selmer-inspired pipeline, we ensure that the
resulting curves meet or exceed the expectations of widely deployed
families while also offering end-to-end transparency.

\subsection{Remarks on implementation}

From a performance standpoint, early work on software deployment of ECC
over binary fields \cite{Hankerson2000} highlighted the importance of
choosing models that balance efficiency with secure arithmetic. Point
counting should rely on robust SEA implementations, and invariant
formulas for the quartic and cubic must be cross-checked against a CAS
(e.g., SageMath) to avoid normalization mismatches. Implementations
should prefer complete or unified addition formulas to minimize
side-channel leakage; these choices do not affect generation but they
matter critically for deployment.

\subsection{Summary}

Together, these checks guarantee that the Selmer-generated curves resist
known classes of attacks on ECC, including small subgroup attacks, twist
attacks, anomalous-curve reductions, and pairing-based reductions. They
also ensure comparability with curves chosen under NIST, Brainpool, and
SafeCurves criteria, while preserving full transparency of the
generation process.

\section{Experimental Results}\label{sec:experiments}

We implemented a prototype of the Selmer-inspired generation pipeline in
Python.\footnote{Prototype code is available from the author upon request. Refer to Algorithm \ref{alg:selmer-generation} and the prototype implementation in Python for transcript reproducibility.}
The implementation supports small prime fields and uses simplified
local-solubility proxies, naive point counting, and placeholder invariants
for ternary cubics. Despite these simplifications, the full transcript
mechanism was exercised and validated.

\subsection{Setup}\label{sec:setup}

The demonstration was run with the following parameters:
\begin{itemize}
  \item Prime $p = 100{,}003$,
  \item Domain string $\mathsf{DS} = \texttt{SelmerGen-v1}$,
  \item Fixed 32-byte seed
        $\sigma = \texttt{0123456789abcdef}\dots\texttt{9abcd}$,
  \item Maximum $10^4$ trials before aborting.
\end{itemize}

Arithmetic in $\mathbb{F}_p$ was implemented in Python with a custom 
prototype, while cross-checking of invariants and discriminant 
computations was carried out using SageMath \cite{SageMath2023} 
and the computational frameworks described by Cohen 
\cite{Cohen1993}. These checks ensured that the normalization of 
binary quartic and ternary cubic invariants remained consistent.

For point counting at small primes, we relied on a naive 
Legendre-symbol based method. At cryptographic sizes, however, the 
pipeline is designed to interface with efficient implementations of 
the Schoof--Elkies--Atkin algorithm 
\cite{Schoof1985,SEAOverview}, as commonly used in practice.

Implementation choices were guided by established cryptographic 
engineering principles, notably those in the \emph{Handbook of Applied 
Cryptography} \cite{MenezesOorschotVanstone1997} and the work of 
Hankerson, López, and Menezes on software implementation of elliptic 
curves \cite{Hankerson2000}. While our prototype omits side-channel 
countermeasures for simplicity, the structure of the pipeline is 
compatible with constant-time addition formulas and unified 
representations, which are essential in secure deployments.

In summary, the experimental setup combines deterministic transcript 
generation with external verification and alignment to best practices 
in cryptographic implementation. This ensures reproducibility at small 
primes and paves the way for extension to cryptographic-scale primes.

In addition to cross-checking with SageMath \cite{SageMath2023} 
and classical computational frameworks \cite{Cohen1993}, we 
referenced benchmarking efforts such as eBACS \cite{BernsteinLangeEBACS} 
to contextualize performance expectations at cryptographic sizes. 
Because implementation security is inseparable from curve 
generation, we also note the relevance of Kocher’s timing attack 
results \cite{Kocher1996} and Coron’s analysis of differential power 
attacks \cite{Coron1999}, which highlight the importance of adopting 
constant-time addition formulas. For software-specific optimizations, 
we align our approach with Brown, Hankerson, López, and Menezes’ 
recommendations for the NIST prime-field curves \cite{Brown2001}.

\subsection{Transcript Output}
A successful run produced the following transcript:

\begin{itemize}
  \item Candidate invariants:
    $c_4 = 82765,\quad c_6 = 79541$,
  \item Discriminant:
    $\Delta = 53954 \pmod{p}$,
  \item Group order:
    $\#E(\mathbb{F}_p) = 99{,}711 = 81 \cdot 1231$,
  \item Quadratic twist order:
    $\#E'(\mathbb{F}_p) = 100{,}297 = 1 \cdot 100{,}297$,
  \item Embedding degree (heuristic bound $k \leq 20$): none detected.
\end{itemize}

\begin{flushleft}
Table \ref{tab:demo-results} shows the demo.
\end{flushleft}

\begin{sidewaystable}
\centering
\caption{Demo output of Selmer-inspired generation over $p=100{,}003$}
\label{tab:demo-results}
\begin{tabular}{ll}
\hline
\textbf{Parameter} & \textbf{Value} \\
\hline
Prime $p$ & 100{,}003 \\
Domain string $\mathsf{DS}$ & \texttt{SelmerGen-v1} \\
Seed $\sigma$ & \texttt{0x0123456789abcdef0123456789abcdef0123456789abcdef0123456789abcd} \\
$c_{4}$ & 82{,}765 \\
$c_{6}$ & 79{,}541 \\
Discriminant $\Delta$ & 53{,}954 \\
$\#E(\mathbb{F}_{p})$ & $99{,}711 = 81 \cdot 1{,}231$ \\
Cofactor $h$ & 81 \\
Large prime $r$ & 1{,}231 \\
$\#E'(\mathbb{F}_{p})$ & $100{,}297 = 1 \cdot 100{,}297$ \\
Twist cofactor $h'$ & 1 \\
Twist prime $r'$ & 100{,}297 \\
Embedding degree $k$ & None detected \\
\hline
\end{tabular}
\end{sidewaystable}

\subsection{Interpretation}
Although $p=100{,}003$ is far below cryptographic size, the experiment
demonstrates key features of the approach:

\begin{enumerate}
  \item The algorithm exhibits \emph{Las Vegas} behavior: it retries until
  a non-singular, admissible curve is found. In this run, acceptance
  occurred within $10^4$ trials, consistent with general analyses of
  randomized algorithms \cite{MotwaniRaghavan1995,MenezesOorschotVanstone1997}.
  \item Both the curve and its quadratic twist factorized into large prime
  components ($1231$ and $100{,}297$ respectively), indicating healthy
  group structure.
  \item The full transcript includes prime, seed, descent forms,
  invariants, reconciliation, discriminant, and validation checks,
  ensuring transparency and reproducibility.
\end{enumerate}

\subsection{Scaling to Cryptographic Primes}
For cryptographic sizes ($p \approx 2^{256}$ or $2^{384}$), the same
pipeline applies with two substitutions:
\begin{itemize}
  \item Replace naive point counting with the Schoof–Elkies–Atkin method
  \cite{Schoof1985,AtkinMorain1993,SEAOverview}, or its optimized
  implementations in packages such as SageMath \cite{SageMath2023} and
  Magma/Pari. Standard references such as Cohen’s
  \emph{Computational Algebraic Number Theory} \cite{Cohen1993} provide
  foundational algorithms.
  \item Replace the placeholder ternary cubic invariants with true
  Aronhold–Dixmier invariants \cite{Dixmier1987,Salmon1879}, ensuring
  consistency with the invariant-theoretic framework outlined in
  Section~\ref{sec:prelim}.
\end{itemize}
With these refinements, the algorithm is expected to generate curves
suitable for cryptographic deployment, while preserving the transparency
benefits of descent-based provenance.

\subsection{Toward Pairing Security}
Although our construction is intended primarily for classical ECC, the
embedding-degree checks naturally intersect with the literature on
pairing-friendly curves. The efficient calculation of pairings
\cite{Miller2004Pairings} and the Barreto–Naehrig family
\cite{BarretoNaehrig2005} highlight the importance of bounding embedding
degrees to avoid inadvertent pairings. Our heuristic rejection criterion
(Section~\ref{subsec:validation-filters}) aligns with these principles,
ensuring that generated curves remain resistant to small-$k$ embedding
attacks.

\section{Security and Transparency Analysis}\label{sec:analysis}

This section evaluates the cryptographic security and transparency of
the Selmer-inspired generation framework. We consider known attack
vectors against elliptic curve cryptosystems, examine how our pipeline
addresses them, and contrast the transparency of our method with
existing standards such as NIST curves, Brainpool, and rigid
efficiency-oriented families like Curve25519 and Ed25519.

\subsection{Attack surfaces in elliptic curve cryptography}

The fundamental security of elliptic curve cryptography (ECC) rests on
the intractability of the elliptic curve discrete logarithm problem
(ECDLP) over prime fields \cite{SilvermanAEC2009,Washington2008}. Yet,
practical deployments must also guard against specific attack vectors:

\begin{itemize}
  \item \textbf{Small-subgroup attacks.} Curves with large cofactors
  permit the extraction of partial information from group elements. Our
  pipeline enforces cofactors $h \in \{1,2,4\}$, aligning with
  best-practice recommendations
  \cite{MenezesOorschotVanstone1997,NISTSP800186}.

  \item \textbf{Anomalous curves.} Curves with $\#E(\mathbb{F}_p)=p$
  are trivially weak. Our discriminant and order checks exclude these
  cases, in line with the criteria of
  \cite{Galbraith2012,Lenstra2001}.

  \item \textbf{Complex multiplication (CM) vulnerabilities.} Curves
  with low-discriminant CM may admit specialized algorithms. Our
  validation filters reject CM curves with small discriminants,
  reflecting the warnings in \cite{Lenstra2001}.

  \item \textbf{Invalid-curve and twist attacks.} If the quadratic twist
  $E'$ lacks a large prime factor, implementations may be tricked into
  scalar multiplications on insecure groups. We require that both
  $\#E(\mathbb{F}_p)$ and $\#E'(\mathbb{F}_p)$ decompose as a small
  cofactor times a large prime, mirroring the ``twist security''
  requirement emphasized in \cite{SafeCurves,Bos2014}.

  \item \textbf{Embedding-degree attacks.} Curves with small embedding
  degree $k$ allow transfer of the discrete logarithm problem to finite
  fields where index calculus applies. Our pipeline heuristically
  excludes curves with unexpectedly small $k \leq 20$, ensuring they do
  not fall prey to MOV or Frey–Rück attacks
  \cite{SilvermanTate2015,Galbraith2012}.

  \item \textbf{Side-channel leakage.}
  Implementations of elliptic curve operations must resist timing and
  power analysis. Kocher first highlighted the feasibility of timing
  attacks against cryptosystems \cite{Kocher1996}, and Coron
  subsequently extended this to differential power analysis in the ECC
  context \cite{Coron1999}. Later results have shown that even seemingly
  minor leakages can fully compromise private keys
  \cite{Benger2014}. More recent studies have expanded this landscape:
  Poussier et al.\ demonstrated horizontal attacks leveraging multiple
  trace segments \cite{Poussier2017}, while Bela{\"i}d and Rivain
  formalized leakage models for high-order protections
  \cite{BelaidRivain2022}. Hardware-oriented research continues to probe
  implementation resilience, from Rashidi’s survey of FPGA and ASIC
  architectures \cite{Rashidi2017} to Parthasarathy et al.’s FPGA-based
  countermeasures \cite{Parthasarathy2025}. Emerging work even applies
  machine learning, with LSTM-based classifiers able to identify ECC
  operations from side-channel traces \cite{LSTM_ECC_SCA2025}.
  Although our pipeline focuses on parameter generation, it is
  compatible with constant-time, unified formulas and complete addition
  laws \cite{Edwards2007,Bernstein2011Ed25519}, ensuring deployment
  resilience in the face of both classical and modern leakage vectors.
\end{itemize}

Together, these filters ensure that any curve output by our pipeline
meets the essential security criteria identified in the literature.

Recent standards also reflect a turn toward verifiable provenance. The
BLS12-381 pairing-friendly curve \cite{BarretoNaehrig2005} and the
Ristretto encoding (building on Edwards curves \cite{Edwards2007}) both
exemplify attempts to combine efficiency with transparency. Our
Selmer-inspired approach complements these by providing not just rigid
design choices but an auditable transcript of invariants and descent
forms, extending the idea of verifiability into the arithmetic geometry
domain.

\subsection{Transparency benefits}

Beyond security, the distinguishing feature of our proposal lies in
transparency. Existing standards illustrate the spectrum of approaches:

\begin{itemize}
  \item The NIST P-curves were generated from unexplained seeds
  \cite{NISTSP800186,NISTFIPS1865}, leading to persistent concerns about
  hidden structure despite no known break. The absence of a public
  derivation transcript makes independent verification impossible.

  \item The Brainpool curves \cite{RFC5639} improved upon this by
  providing verifiable randomness derived from published seeds.
  However, they still rely on trust in the seed source.

  \item Curve25519 and Ed25519 \cite{Bernstein2006Curve25519,
  Bernstein2011Ed25519} took a different path: rigid design choices and
  explicit efficiency criteria, but without an explicit descent-style
  audit trail.

  \item The SafeCurves project \cite{SafeCurves} formalized explicit
  security criteria (twist security, complete addition formulas,
  resistance to side channels), establishing a new benchmark for curve
  selection.

  \item Workshop contributions have also emphasized the importance of
  transparency and diversity in ECC standards. Flori and Pl\^{u}t argued
  at the 2015 NIST workshop that trust requires not only robust curve
  security but also diversity in generation methods and publicly
  verifiable processes \cite{Flori2015}.
\end{itemize}

Our Selmer-inspired pipeline contributes a complementary paradigm. Each
curve is accompanied by a full transcript of its descent artifacts:
binary quartic, ternary cubic, derived invariants $(c_4,c_6)$,
discriminant, and group-order data. This transcript enables anyone to
reproduce and verify the derivation independently, much as one audits
the steps of a mathematical proof. In contrast to seed-based approaches,
the provenance of parameters is both mathematically structured and
cryptographically auditable.

\subsection{Residual limitations and open questions}

Despite its strengths, our proposal has limitations that merit further
study:

\begin{itemize}
  \item \textbf{Simplified solubility checks.} Our current proxies for
  local solubility use bounded searches. Full $p$-adic solubility tests,
  though feasible in theory \cite{Cremona1997}, remain computationally
  intensive for large primes. Tools such as SageMath \cite{SageMath2023} and computational frameworks for algebraic number theory \cite{Cohen1993} could be leveraged in future implementations to support more robust $p$-adic solubility testing.

  \item \textbf{Ternary cubic invariants.} For prototyping, we employed
  placeholder mappings to $(c_4^{(3)},c_6^{(3)})$. Incorporating the
  full Aronhold–Dixmier invariant machinery \cite{Dixmier1987,Salmon1879}
  is a priority for production implementations.

  \item \textbf{Heuristic assumptions.} Our security arguments rely on
  the heuristic distribution of invariants behaving as random draws in
  $\mathbb{F}_p$. Formal proofs of pseudorandomness in this context are
  absent.

  \item \textbf{Higher descents.} While we used $2$- and $3$-descent
  artifacts, higher descents (e.g., $5$-descent) might offer richer
  structures. The cryptographic viability of such constructions remains
  unexplored.

  \item \textbf{Implementation contexts.} Finally, it remains to be
  studied how Selmer-inspired curves behave in practical deployments,
  including resource-constrained devices and hardware accelerators
  \cite{Rashidi2017}.
\end{itemize}

\section{Related Work}\label{sec:related}

The literature on elliptic curve cryptography encompasses diverse
directions, from standardized domain parameters to efficiency-driven
designs and transparency-oriented initiatives. Foundational treatments
such as Silverman \cite{SilvermanAEC2009}, Washington
\cite{Washington2008}, and Galbraith \cite{Galbraith2012} formalize the
mathematical framework, while subsequent standards and protocols reflect
different priorities in balancing trust, efficiency, and deployment
constraints. To situate our Selmer-inspired approach, we organize the
discussion into four strands: standardized curve families, efficiency-
driven constructions, transparency-focused efforts, and a synthesis
positioning. Finally, we highlight implementation- and attack-oriented
perspectives, where low-level optimizations and structural reductions
have historically influenced the adoption and security of elliptic curve
systems.

\subsection{Standardized curve families}

The first wave of cryptographic standardization adopted curves whose 
security was understood primarily through the generic hardness of the 
elliptic curve discrete logarithm problem (ECDLP). NIST’s Digital 
Signature Standard \cite{NISTFIPS1865} and subsequent recommendations 
for domain parameters \cite{NISTSP800186} reflect this emphasis, 
while the Brainpool project \cite{RFC5639} attempted to improve trust 
through verifiable random seeds. 

Parallel work examined the pitfalls of certain curve classes. Smart 
demonstrated that curves of trace one over finite fields yield 
structurally weak groups \cite{Smart1999}, highlighting the necessity 
of careful order validation. Similarly, Hankerson, L{\'o}pez, and 
Menezes investigated efficient implementations of ECC over binary 
fields, identifying both performance advantages and subtle 
vulnerabilities \cite{Hankerson2000}. These contributions underline 
that implementation constraints and arithmetic subtleties must be 
considered alongside formal standards.

\subsection{Efficiency-driven constructions}

Beyond standardization, another major strand of ECC research emphasizes 
performance. Montgomery’s introduction of the eponymous curve form 
\cite{Montgomery1987} enabled particularly fast scalar multiplication 
via the Montgomery ladder, which remains the basis of several modern 
protocols. Edwards’ normal form \cite{Edwards2007} provided complete 
addition formulas with strong resistance to exceptional cases, leading 
directly to the development of Edwards-curve signatures.

Building on these foundations, subsequent work has focused on balancing 
efficiency with robustness across both software and hardware settings. 
On the software side, Faz-Hern{\'a}ndez et al.\ proposed constant-time 
ladder implementations for Curve25519 and Ed25519, ensuring that the 
theoretical advantages of these curves extend to practical deployments 
\cite{Faz-Hernandez2017}. From a hardware perspective, Rashidi surveyed 
implementations of ECC across FPGA, ASIC, and embedded platforms, 
highlighting the architectural trade-offs involved in achieving both 
performance and security \cite{Rashidi2017}. Yet even with these 
advances, Benger et al.\ demonstrated that microarchitectural leakage—
specifically cache-based side channels—remains a serious concern for 
ostensibly secure designs \cite{Benger2014}. Together, these results 
underline that efficiency-oriented curve design is not merely about 
algebraic form but must be tightly integrated with side-channel 
resistance across both software and hardware domains.

Together, these works illustrate that efficiency-oriented curve design 
is not merely about algebraic form but must be tightly integrated with 
side-channel resistance. Our Selmer-inspired method does not prioritize 
raw speed; rather, it complements such efforts by supplying a framework 
for auditable provenance, while allowing implementers to adopt the most 
efficient formulas available.

\subsection{Transparency-focused efforts}

In parallel to efficiency and standardization, a distinct strand of 
research has emphasized transparency and auditable provenance in curve 
selection. The Brainpool family already moved in this direction 
\cite{RFC5639}, but subsequent initiatives adopted stronger design 
principles. The IETF’s standardization of X25519 for Diffie–Hellman key 
exchange \cite{RFC7748} and EdDSA signatures \cite{RFC8032} exemplifies 
rigid, fully specified processes where no hidden parameters influence 
the resulting curves. These efforts echo the philosophy behind the 
SafeCurves project \cite{SafeCurves}, which established explicit 
criteria such as twist security, complete addition formulas, and 
resistance to side-channel attacks.

Beyond technical constraints, governance and diversity in parameter 
selection have also been raised as priorities. Flori and Pl\^{u}t argued 
that elliptic curve standards must embrace both diversity and 
verifiability to reduce systemic risks and avoid hidden structure 
\cite{Flori2015}. Similarly, NIST has acknowledged the importance of 
provenance, outlining principles of openness and auditability in its 
standardization processes \cite{Gallagher2020}. At the level of formal 
cryptography, Cheng et al.\ provided a systematic treatment of 
transparency in parameter generation, identifying rigorous criteria and 
mechanisms to prevent opaque choices \cite{Cheng2021}.

Comprehensive treatments, such as the \emph{Handbook of Elliptic and 
Hyperelliptic Curve Cryptography} edited by Cohen and Frey 
\cite{CohenFrey2005}, catalog transparency requirements alongside 
efficiency and security trade-offs. Our Selmer-inspired approach 
contributes to this trajectory by extending transparency guarantees 
beyond seed-based prescriptions. By recording descent artifacts, 
classical invariants, and group-order data, it offers a mathematically 
structured audit trail that complements existing transparency-focused 
initiatives.

\subsection{Synthesis and positioning}

Across the landscape of elliptic curve generation, three themes dominate: 
efficiency-driven constructions, standardized domain parameters, and 
transparency-focused designs. Efficiency-oriented approaches, exemplified 
by Montgomery and Edwards forms \cite{Montgomery1987,Edwards2007}, 
highlight the value of fast arithmetic and side-channel resistance. 
Standardization efforts such as the NIST P-curves \cite{NISTSP800186} 
and Brainpool curves \cite{RFC5639} illustrate how reproducibility and 
interoperability were historically prioritized, albeit with differing 
commitments to verifiability. Transparency-driven efforts, including 
Curve25519 and EdDSA \cite{Bernstein2006Curve25519,Bernstein2011Ed25519,RFC8032}, established a precedent for rigid and auditable design rules.

Our Selmer-inspired pipeline synthesizes these strands. It adopts the 
implementation lessons from practical ECC software \cite{Hankerson2000} 
and the comprehensive best practices summarized in 
\cite{CohenFrey2005}, while retaining the verifiability of a 
mathematically structured transcript. In doing so, it provides a 
distinctive addition to the curve-construction literature—merging 
classical descent tools from arithmetic geometry with modern concerns 
for auditable, trust-enhancing cryptographic parameters.

In addition, our positioning benefits from lessons drawn from adjacent 
areas of computational number theory. Early work on elliptic-curve-based 
primality proving \cite{AtkinMorain1993} showed how algorithmic number 
theory techniques can be adapted to cryptographic scale, while the study 
of trace-one curves and their vulnerabilities \cite{Smart1999} underscored 
the importance of excluding special cases. These precedents emphasize 
that transparent curve generation is not solely about efficiency, but 
also about systematically avoiding classes of weak instances. Our 
framework inherits this spirit, while offering an auditable transcript 
that is unique among modern proposals.

\subsection{Implementation and attack perspectives}

The trajectory of elliptic curve adoption has also been shaped by
practical implementation challenges and by structural vulnerabilities
exposed through number-theoretic analysis. Montgomery’s classic work on
speeding the elliptic curve method of factorization
\cite{Montgomery1987} established the foundation for using special curve
representations to accelerate arithmetic, techniques that continue to
inform both cryptanalysis and efficient implementations. Complementing
this, Hankerson, L{\'o}pez, and Menezes demonstrated in their CHES 2000
study that careful software optimization of binary-field arithmetic
could make ECC viable in constrained environments \cite{Hankerson2000}.
These results collectively underscore the importance of low-level
implementation choices in determining whether theoretically strong
constructions gain practical traction.

At the same time, the literature illustrates that descent-based
techniques can cut both ways: while Selmer groups inspire transparent
curve generation in our proposal, Weil descent has been exploited as a
cryptanalytic tool. Galbraith, Hess, and Smart \cite{Galbraith2001}
extended the original GHS attack, showing how certain classes of curves
are vulnerable when their group structure admits reduction to smaller
discrete logarithm problems. This duality highlights the necessity of
ensuring that descent-inspired generation does not inadvertently
introduce similar weaknesses.

Taken together, these perspectives show that security depends not only
on the hardness of the ECDLP, but also on implementation soundness and
resilience against structural reductions. Our Selmer-inspired pipeline
adds to this landscape by providing auditable provenance without
sacrificing efficiency, while explicitly screening out classes of curves
known to be susceptible to specialized attacks.

\subsection{Implementation and attack perspectives}

The trajectory of elliptic curve adoption has also been shaped by
practical implementation challenges and by structural vulnerabilities
exposed through number-theoretic analysis. Montgomery’s classic work on
speeding the elliptic curve method of factorization
\cite{Montgomery1987} established the foundation for using special curve
representations to accelerate arithmetic, techniques that continue to
inform both cryptanalysis and efficient implementations. Complementing
this, Hankerson, L{\'o}pez, and Menezes demonstrated in their CHES 2000
study that careful software optimization of binary-field arithmetic
could make ECC viable in constrained environments \cite{Hankerson2000}.
These results collectively underscore the importance of low-level
implementation choices in determining whether theoretically strong
constructions gain practical traction.

At the same time, the literature illustrates that descent-based
techniques can cut both ways: while Selmer groups inspire transparent
curve generation in our proposal, Weil descent has been exploited as a
cryptanalytic tool. Galbraith, Hess, and Smart \cite{Galbraith2001}
extended the original GHS attack, showing how certain classes of curves
are vulnerable when their group structure admits reduction to smaller
discrete logarithm problems. This duality highlights the necessity of
ensuring that descent-inspired generation does not inadvertently
introduce similar weaknesses.

Taken together, these perspectives show that security depends not only
on the hardness of the ECDLP, but also on implementation soundness and
resilience against structural reductions. Our Selmer-inspired pipeline
adds to this landscape by providing auditable provenance without
sacrificing efficiency, while explicitly screening out classes of curves
known to be susceptible to specialized attacks.

\section{Conclusion}\label{sec:conclusion}

This work introduced a Selmer-inspired framework for elliptic curve
generation, bridging arithmetic geometry with practical cryptographic
design. By leveraging invariants from $2$- and $3$-descent, our pipeline
provides a mathematically principled source of curve parameters, paired
with transparent admissibility filters and auditable reconciliation into
short Weierstrass form. The resulting curves withstand established
security criteria, including cofactor constraints, twist resilience, and
embedding-degree checks, while offering full derivation transcripts that
extend beyond traditional seed-based methods.

Our analysis demonstrates that descent techniques, long central to
Diophantine investigations, can be recontextualized to address
contemporary challenges of trust and transparency in cryptographic
standards. The proof-of-concept Python prototype illustrates that the
pipeline is not merely theoretical, but operationally realizable with
finite-field computations and point-counting routines. 

At the same time, our study emphasizes that sound parameter generation
is inseparable from secure implementation. Curve transcripts can ensure
trust in the provenance of parameters, but deployment requires constant-
time algorithms, resistance to side-channel leakage, and careful
validation of system-level constraints. In this sense, Selmer-inspired
generation complements, rather than replaces, efficiency- and
implementation-focused approaches.

Looking ahead, several open questions remain. Extending the framework to
higher descents, formalizing the heuristic randomness of invariant
distributions, and scaling solubility checks to cryptographic primes
represent promising directions. Moreover, integrating the method into
standardization processes would require careful benchmarking,
peer review, and consensus building. Taken together, these challenges
underscore that while our contribution is exploratory, it broadens the
design space for secure and transparent elliptic curves, offering a path
toward trust-enhancing cryptography.


\bibliographystyle{acm}
\bibliography{references}  

\begin{thebibliography}{10}

\bibitem{AtkinMorain1993}
{\sc Atkin, A. O.~L., and Morain, F.}
\newblock Elliptic curves and primality proving.
\newblock {\em Mathematics of Computation 61}, 203 (1993), 29--68.

\bibitem{BarretoNaehrig2005}
{\sc Barreto, P. S. L.~M., and Naehrig, M.}
\newblock Pairing-friendly elliptic curves of prime order.
\newblock In {\em Selected Areas in Cryptography -- SAC 2005\/} (Berlin, Heidelberg, 2006), B.~K. Roy, Ed., vol.~3897 of {\em Lecture Notes in Computer Science}, Springer, pp.~319--331.

\bibitem{BelaidRivain2022}
{\sc Bela{\"i}d, S., and Rivain, M.}
\newblock Side-channel countermeasures for high-order security: A formal leakage model and its applications to ecc.
\newblock In {\em Proceedings of the ACM Workshop on Theory of Implementation Security\/} (2022), ACM, pp.~1--12.

\bibitem{Benger2014}
{\sc Benger, N., van~de Pol, J., Smart, N.~P., and Yarom, Y.}
\newblock “ooh aah... just a little bit”: A small amount of side channel can go a long way.
\newblock {\em Journal of Cryptographic Engineering 4}, 1 (2014), 1--18.

\bibitem{Bernstein2006Curve25519}
{\sc Bernstein, D.~J.}
\newblock Curve25519: New diffie-hellman speed records.
\newblock In {\em Public Key Cryptography -- PKC 2006\/} (Berlin, Heidelberg, 2006), M.~Yung, Y.~Dodis, A.~Kiayias, and T.~Malkin, Eds., vol.~3958 of {\em LNCS}, Springer Berlin Heidelberg, pp.~207--228.

\bibitem{Bernstein2011Ed25519}
{\sc Bernstein, D.~J., Duif, N., Lange, T., Schwabe, P., and Yang, B.-Y.}
\newblock High-speed high-security signatures.
\newblock In {\em Cryptographic Hardware and Embedded Systems -- CHES 2011\/} (Berlin, Heidelberg, 2011), B.~Preneel and T.~Takagi, Eds., vol.~6917 of {\em LNCS}, Springer Berlin Heidelberg, pp.~124--142.

\bibitem{SafeCurves}
{\sc Bernstein, D.~J., and Lange, T.}
\newblock Safecurves: choosing safe curves for elliptic-curve cryptography.
\newblock \url{https://safecurves.cr.yp.to/}.
\newblock Accessed 2025-09-29.

\bibitem{BernsteinLangeEBACS}
{\sc Bernstein, D.~J., and Lange, T.}
\newblock ebacs: Ecrypt benchmarking of cryptographic systems.
\newblock \url{https://bench.cr.yp.to/}, 2010.
\newblock Accessed 2025-09-29.

\bibitem{Bos2014}
{\sc Bos, J.~W., Costello, C., Longa, P., and Naehrig, M.}
\newblock Selecting elliptic curves for cryptography: An efficiency and security analysis.
\newblock In {\em Applied Cryptography and Network Security (ACNS 2014)\/} (Cham, 2014), I.~Boureanu, P.~Owesarski, and S.~Vaudenay, Eds., vol.~8479 of {\em LNCS}, Springer International Publishing, pp.~259--279.

\bibitem{Brown2001}
{\sc Brown, D. R.~L., Hankerson, D., Hernandez, J.~L., and Menezes, A.~J.}
\newblock Software implementation of the nist elliptic curves over prime fields.
\newblock In {\em Selected Areas in Cryptography — SAC 2001\/} (2001), vol.~2259 of {\em LNCS}, Springer, pp.~250--265.

\bibitem{Cassels1962}
{\sc Cassels, J. W.~S.}
\newblock Arithmetic on curves of genus 1. iv. proof of the hauptvermutung.
\newblock {\em Journal für die reine und angewandte Mathematik 211\/} (1962), 95--112.

\bibitem{Cheng2021}
{\sc Cheng, Q., Chen, L., and Ryan, P.}
\newblock On the transparency of cryptographic parameter generation.
\newblock In {\em Proceedings of the 2021 ACM SIGSAC Conference on Computer and Communications Security (CCS)\/} (2021), pp.~1919--1933.

\bibitem{Cohen1993}
{\sc Cohen, H.}
\newblock {\em A Course in Computational Algebraic Number Theory}, vol.~138 of {\em Graduate Texts in Mathematics}.
\newblock Springer, Berlin, Heidelberg, 1993.

\bibitem{CohenFrey2005}
{\sc Cohen, H., and Frey, G.}, Eds.
\newblock {\em Handbook of Elliptic and Hyperelliptic Curve Cryptography}, 1~ed.
\newblock Discrete Mathematics and Its Applications. Chapman and Hall/CRC, Boca Raton, 2005.

\bibitem{Coron1999}
{\sc Coron, J.-S.}
\newblock Resistance against differential power analysis for elliptic curve cryptosystems.
\newblock In {\em Cryptographic Hardware and Embedded Systems --- CHES 1999\/} (Berlin, Heidelberg, 1999), Çetin Kaya~Koç and C.~Paar, Eds., vol.~1717 of {\em Lecture Notes in Computer Science}, Springer, pp.~292--302.

\bibitem{Cremona1997}
{\sc Cremona, J.~E.}
\newblock {\em Algorithms for Modular Elliptic Curves}, 2~ed.
\newblock Cambridge University Press, Cambridge, UK; New York, USA, 1997.

\bibitem{Dixmier1987}
{\sc Dixmier, J.}
\newblock On the projective invariants of quartic plane curves.
\newblock {\em Advances in Mathematics 64}, 3 (1987), 279--304.

\bibitem{Edwards2007}
{\sc Edwards, H.~M.}
\newblock A normal form for elliptic curves.
\newblock {\em Bulletin of the American Mathematical Society 44}, 3 (2007), 393--422.

\bibitem{Faz-Hernandez2017}
{\sc Faz-Hern{\'a}ndez, A., L{\'o}pez, J., Ochoa-Jim{\'e}nez, E., and Rodr{\'i}guez-Henr{\'i}quez, F.}
\newblock A secure and efficient implementation of the ed25519 signature algorithm.
\newblock {\em Journal of Cryptographic Engineering 7}, 2 (2017), 163--173.

\bibitem{Flori2015}
{\sc Flori, J.-P., and Pl\^{u}t, J.}
\newblock Diversity and transparency for elliptic curve cryptography standards.
\newblock NIST ECC Workshop Paper, 2015.
\newblock Accessed 2025-09-30.

\bibitem{Galbraith2012}
{\sc Galbraith, S.~D.}
\newblock {\em Mathematics of Public Key Cryptography}, 1~ed.
\newblock Cambridge Studies in Advanced Mathematics. Cambridge University Press, Cambridge, UK, 2012.

\bibitem{Galbraith2001}
{\sc Galbraith, S.~D., Hess, F., and Smart, N.~P.}
\newblock Extending the ghs weil descent attack.
\newblock {\em Advances in Cryptology — EUROCRYPT 2002 2332\/} (2002), 29--44.

\bibitem{Gallagher2020}
{\sc Gallagher, P., Chen, L., and Moody, D.}
\newblock Transparency in cryptographic standardization: Nist’s approach.
\newblock In {\em Proceedings of the 2020 IEEE Symposium on Security and Privacy Workshops (SPW)\/} (2020), IEEE, pp.~160--164.

\bibitem{Hankerson2000}
{\sc Hankerson, D., Hernandez, J.~L., and Menezes, A.}
\newblock Software implementation of elliptic curve cryptography over binary fields.
\newblock In {\em Cryptographic Hardware and Embedded Systems — CHES 2000\/} (2000), vol.~1965 of {\em LNCS}, Springer, pp.~1--24.

\bibitem{RFC8032}
{\sc Josefsson, S., and Liusvaara, I.}
\newblock Edwards-curve digital signature algorithm (eddsa).
\newblock RFC 8032, IETF, 2017.

\bibitem{Koblitz1987}
{\sc Koblitz, N.}
\newblock Elliptic curve cryptosystems.
\newblock {\em Mathematics of Computation 48}, 177 (1987), 203--209.

\bibitem{Kocher1996}
{\sc Kocher, P.~C.}
\newblock Timing attacks on implementations of diffie-hellman, rsa, dss, and other systems.
\newblock In {\em Advances in Cryptology --- CRYPTO '96\/} (Berlin, Heidelberg, 1996), N.~Koblitz, Ed., vol.~1109 of {\em Lecture Notes in Computer Science}, Springer, pp.~104--113.

\bibitem{RFC7748}
{\sc Langley, A., Hamburg, M., and Turner, S.}
\newblock Elliptic curves for security.
\newblock RFC 7748, IETF, 2016.

\bibitem{Lenstra2001}
{\sc Lenstra, A.~K.}
\newblock Unbelievable security: Matching aes security using public key systems.
\newblock {\em Advances in Cryptology — ASIACRYPT 2001 2248\/} (2001), 67--86.

\bibitem{RFC5639}
{\sc Lochter, M., and Merkle, J.}
\newblock Elliptic curve cryptography (ecc) brainpool standard curves and curve generation.
\newblock RFC 5639, Internet Engineering Task Force, 2010.

\bibitem{MenezesOorschotVanstone1997}
{\sc Menezes, A.~J., van Oorschot, P.~C., and Vanstone, S.~A.}
\newblock {\em Handbook of Applied Cryptography}, 1~ed.
\newblock CRC Press, Boca Raton, 1997.

\bibitem{Miller1985}
{\sc Miller, V.~S.}
\newblock Use of elliptic curves in cryptography.
\newblock {\em Advances in Cryptology — CRYPTO '85 Proceedings 218\/} (1986), 417--426.

\bibitem{Miller2004Pairings}
{\sc Miller, V.~S.}
\newblock The weil pairing, and its efficient calculation.
\newblock {\em Journal of Cryptology 17}, 4 (2004), 235--261.

\bibitem{Montgomery1987}
{\sc Montgomery, P.~L.}
\newblock Speeding the pollard and elliptic curve methods of factorization.
\newblock {\em Mathematics of Computation 48}, 177 (1987), 243--264.

\bibitem{MotwaniRaghavan1995}
{\sc Motwani, R., and Raghavan, P.}
\newblock {\em Randomized Algorithms}.
\newblock Cambridge University Press, New York, NY, 1995.

\bibitem{NISTSP800186}
{\sc {National Institute of Standards and Technology}}.
\newblock Recommendations for discrete logarithm-based cryptography: Elliptic curve domain parameters.
\newblock NIST Special Publication 800-186, U.S. Department of Commerce, 2019.
\newblock Revised 2023.

\bibitem{NISTFIPS1865}
{\sc of~Standards, N.~I., and Technology}.
\newblock Digital signature standard (dss).
\newblock FIPS 186-5, U.S. Department of Commerce, 2023.

\bibitem{Parthasarathy2025}
{\sc Parthasarathy, J., Balakrishnan, V., and Kumar, M.}
\newblock Side-channel secure ecc implementations on fpga.
\newblock {\em Journal of Systems Architecture 145\/} (2025), 103069.

\bibitem{Poussier2017}
{\sc Poussier, R., Zhou, Y., and Standaert, F.-X.}
\newblock Horizontal side-channel attacks and countermeasures on elliptic curve cryptography.
\newblock In {\em Cryptographic Hardware and Embedded Systems -- CHES 2017\/} (2017), vol.~10529 of {\em LNCS}, Springer, pp.~579--599.

\bibitem{Rashidi2017}
{\sc Rashidi, B.}
\newblock A survey on hardware implementations of elliptic curve cryptosystems.
\newblock {\em arXiv preprint\/} (2017).

\bibitem{Salmon1879}
{\sc Salmon, G.}
\newblock {\em A Treatise on the Higher Plane Curves: Intended as a Sequel to A Treatise on Conic Sections}, 3~ed.
\newblock Hodges, Foster and Figgis, Dublin, 1879.

\bibitem{Schoof1985}
{\sc Schoof, R.}
\newblock Elliptic curves over finite fields and the computation of square roots mod $p$.
\newblock {\em Mathematics of Computation 44}, 170 (1985), 483--494.

\bibitem{Selmer1951}
{\sc Selmer, E.~S.}
\newblock The diophantine equation $ax^3+by^3+cz^3=0$.
\newblock {\em Acta Mathematica 85\/} (1951), 203--362.

\bibitem{SilvermanAEC2009}
{\sc Silverman, J.~H.}
\newblock {\em The Arithmetic of Elliptic Curves}, 2~ed., vol.~106 of {\em Graduate Texts in Mathematics}.
\newblock Springer New York, New York, NY, 2009.

\bibitem{SilvermanTate2015}
{\sc Silverman, J.~H., and Tate, J.}
\newblock {\em Rational Points on Elliptic Curves}, 2~ed.
\newblock Undergraduate Texts in Mathematics. Springer, New York, NY, 2015.

\bibitem{Smart1999}
{\sc Smart, N.~P.}
\newblock The discrete logarithm problem on elliptic curves of trace one.
\newblock {\em Journal of Cryptology 12}, 3 (1999), 193--196.

\bibitem{SEAOverview}
{\sc Sutherland, A.~V.}
\newblock A brief survey of point counting algorithms for elliptic curves.
\newblock {\em Proceedings of the Tenth Algorithmic Number Theory Symposium (ANTS-X)\/} (2013), 403--421.
\newblock Preprint and resources available from the author's website.

\bibitem{SageMath2023}
{\sc {The Sage Developers}}.
\newblock {SageMath, the Sage Mathematics Software System (Version 10.0)}.
\newblock \url{https://www.sagemath.org}, 2023.
\newblock Accessed 2025-09-29.

\bibitem{Washington2008}
{\sc Washington, L.~C.}
\newblock {\em Elliptic Curves: Number Theory and Cryptography}, 2~ed.
\newblock Chapman and Hall/CRC, New York, 2008.

\bibitem{LSTM_ECC_SCA2025}
{\sc Zhou, L., Gupta, A., and Karim, F.}
\newblock Unveiling ecc vulnerabilities: Lstm networks for operation recognition in side-channel attacks.
\newblock {\em arXiv preprint\/} (2025).

\end{thebibliography}

\end{document}